\documentclass[conference]{IEEEtran}

\usepackage{graphicx}
\usepackage{color}
\usepackage[utf8]{inputenc}
\usepackage{enumerate}
\usepackage{amsmath}
\usepackage{amsthm}
\usepackage{amssymb}
\usepackage{accents}
\usepackage{subfig}
\usepackage[style=ieee,backend=bibtex]{biblatex}
\usepackage{algorithm}
\usepackage{algorithmic}
\usepackage{eurosym}
\usepackage{todonotes}
\usepackage{soul} 

\newcommand{\starred}[1]{\accentset{\star}{#1}}

\newcommand{\us}[1]{\overline{#1}}
\newcommand{\ls}[1]{\underline{#1}}

\newcommand{\fc}[1]{\hat{#1}}

\title{A quantitative analysis of the effect of \\flexible loads on reserve markets}
\author{S. Mathieu\footnote{Corresponding author.}, Q. Louveaux, D. Ernst, B. Corn\'elusse\\
 Department of Electrical Engineering and Computer Science,\\
  University of Li\`ege, Li\`ege 4000, Belgium\\
\{sebastien.mathieu,bertrand.cornelusse,dernst,q.louveaux\}@ulg.ac.be}
\date{\today}

\begin{document}
\maketitle
\begin{abstract}
We propose and analyze a day-ahead reserve market model that handles bids from flexible loads. 
This pool market model takes into account the fact that a load modulation in one direction must usually  be compensated later by a modulation of the same magnitude in the opposite direction. 
Our analysis takes into account the gaming possibilities of producers and retailers, controlling load flexibility, in the day-ahead energy and reserve markets, and in imbalance settlement. 
This analysis is carried out by an agent-based approach where, for every round, each actor uses linear programs to maximize its profit according to forecasts of the prices. 
The procurement of a reserve is assumed to be determined, for each period, as a fixed percentage of the total consumption cleared in the energy market for the same period. 
The results show that the provision of reserves by flexible loads has a negligible impact on the energy market prices but markedly decreases the cost of reserve procurement.
However, as the rate of flexible loads increases, the system operator has to rely more and more on non-contracted reserves, which may cancel out the benefits made in the procurement of reserves. 
\end{abstract}

\IEEEpeerreviewmaketitle

\section{Introduction}
\label{sec:introduction}

\IEEEPARstart{W}{ith} the rapid development of renewable energy, the different actors in the electrical power sector are increasingly trying to exploit load flexibility. 
For example, many retailers are in the process of developing the required infrastructure, either directly or through a separate entity called an aggregator, for moving consumption to periods of the day for which the price of electricity is low. 
Another example is the case of System Operators (SOs) that are increasingly trying to attract flexible loads to their market for ancillary services, by allowing, for example, loads at the distribution level to take part in it.

Many authors have studied the effects of load flexibility in the electricity and reserve market on the payoffs to different actors in the electrical system \cite{Artac2012Flexible,wang2003demand, karangelos2012towards}. 
In \cite{Artac2012Flexible}, demand is represented by constant elasticity curves independent from one market period to another.
The results show that demand-side reserve provision leads to lower operating costs.
Reference \cite{wang2003demand} proposes a market model where the demand-side is directly controlled so as to shift consumption and provide upward and downward reserves. 
They also conclude that demand-side reserve offers can lead to significant gains in economic efficiency.
Load reduction periods are typically followed by load recovery periods \cite{karangelos2012towards}.
This observation leads the author to conclude that first, the demand-side should not be seen as a pure alternative to the provision of reserves, and 
second, the participation of demand-side resources could increase the overall required levels of reserve.
Nevertheless the fact that the system operator can exploit demand flexibility can reduce operating costs.
These results are based on globally optimized systems and do not capture gaming possibilities coming from the individual optimization of each actor.

This paper aims to capture these gaming possibilities in an agent-based model.
First, we model the behavior of the actors (producers and retailers acting on load flexibility) of the electricity market and the market mechanisms governed by the system operator.
Afterwards, the models are used to simulate the electrical market system.
Finally, the trajectories of the system are analyzed so as to provide answers concerning the properties of the electricity market. 
This paper is a first step towards answering questions such as the following.
How is the money paid for reserve procurement affected by allowing more flexible loads to participate in the reserve market? 
How would it affect energy market prices? 
How would a retailer behave when it could exploit its load flexibility both in the energy market and the reserve market?

Agent-based modeling has been extensively applied to electricity markets \cite{ventosa2005electricity}. 
The review \cite{weidlich2008critical} concludes that most models represent the demand side as a fixed and price-insensitive load.
For instance, \cite{bunn2003evaluating} models the electricity trading arrangements in the United Kingdom. 
In this system, retailers supply inelastic loads but may game on an intentional imbalance to maximize their profit.
Nowadays, retailers already have access to flexible loads to optimize their costs. 
In this paper, we assume that retailers have direct control over the flexible loads in their portfolio.
Control with real-time pricing is investigated in \cite{Zugno2013182}, 
where retailers optimize the real-time pricing to minimize their retailing and imbalance costs.
The results show that a retailer has an incentive to shift the demand
using a time-dependent price to reduce its imbalance.
These models do not consider the provision of secondary reserves by load aggregation.

In this paper, we make several assumptions concerning the energy market and the reserve market.
The methodology used in this paper could be applied straightforwardly to alternative hypotheses.
First, we assume that both markets are day-ahead markets. 
Second, we suppose that they are cleared sequentially. 
Third, we assume that these markets are pool markets where the market prices are the price for energy and the price for flexibility. 
Our simulation is based on three main stages. In the first stage, producers and retailers submit their bids to the energy market for each period of the next day. 
In the second stage, bids are submitted to the reserve market. 
In a first setting, we only allow producers to submit bids whereas in a second setting, both producers and retailers are allowed to bid in the secondary reserve market.
Finally, the third stage gives the imbalance fee that each actor pays or receives depending on its position at the end of each market period compared to the position announced one day ahead.

For the sake of simplicity, we consider only three types of actors: retailers with flexible demand in their portfolio, producers of energy, and the SO which has to buy in the retail market well-defined amounts of flexibility. 
We make several assumptions to model the behavior of the other actors. 
The main one is that they optimize their positions using forecasts of the prices of the energy and reserve markets, and the imbalance tariffs.
In the simulations described in this paper, we arbitrarily suppose that they use a weighted average of the previous prices to forecast prices.

The rest of the paper is organized as follows. 
Section \ref{sec:currentMarket} focuses on the case where flexible loads can only be exploited in the energy market.
It defines the models that are used by the actors, and presents simulation results that serve for comparison with the results of the next section.
 Section~\ref{sec:market2} proposes a reorganization of the system to allow retailers to bid in the secondary reserve market, and computational results are compared to the first setting.
Section \ref{sec:conclusion} concludes.



\section{Model of the current system}
\label{sec:currentMarket}
This section presents the current organization of the system.
It details the mathematical problems that each stakeholder solves to optimize its decisions.
Stakeholders have three decision stages, which are summarized here, and detailed in the next subsections. 
One day ahead, the energy market, through market operators, collects the offers of the participants (in our case producers and retailers), computes a uniform price for each period, and notifies the participants of the acceptance of their offers. 
In a second stage, still one day ahead but after the clearing of the energy market, producers can bid in the secondary reserve market. 
After these two stages, producers and retailers optimize their position according to an estimate of the imbalance tariffs for the next day, and submit to the SO the net power they will inject/withdraw from the network in their balancing perimeter. 
The participants pay or receive an imbalance fee depending on their position at the end of each market period compared to the position they announced one day ahead.

The main simplifications we make are the following. First, we assume each type of actor makes decisions according to the same mathematical
model, but with its own data, and solves these mathematical models to optimality. 
Second, we use linear programming relaxations of the problems that are actually solved by the stakeholders.

\subsection{Energy market}
The energy market model takes its inspiration from the Central Western Europe coupled market \cite{COSMOS}. 
Producers and retailers submit offers to the market operator. 
An offer is defined by a volume and a limit price, and can span only one time period $t \in \{1,\ldots,T\}$. 
The proportion of an offer which is accepted is determined by the market clearing price (MCP) computed by the market operator.
The MCP is the price at the intersection of the supply and the demand curves, which depends on the period, and is denoted by $\pi^E_t$. 
Supply offers are fully rejected (resp. accepted) when their limit price is greater (resp. smaller) than the MCP, or partially accepted when their limit price equals the MCP. 
The MCP is bounded by $\pi^{cap}$.

For each period $t$, each actor forecasts the price of the energy market $\fc{\pi}^E_t$. This forecast is obtained by the exponential mean of the prices in the last $T$ rounds.
Note that if the price cap $\pi^{cap}$ is reached for a round in the history, the value is replaced in the mean by the last non-capped one. 
However, the information that the price cap was reached in a round is directly integrated into the optimization model of the actors.

\subsection{Reserve market}
\label{sec:currentReserveMarket}
We assume that the SO has to procure a quantity of upward and downward reserve for each period, respectively $R^+_t$ and $R^-_t$, determined as a fixed percentage of the total consumption in period $t$. A bid in the reserve market consists of a maximum power (positive for upward reserve, negative otherwise), and an activation price.
In case of activation, the SO pays the activation price times the energy activated.
Every bid covers only one period and may be rejected, accepted partially, or totally.
The reservation of the capacity of a bid is remunerated at a regulated  capacity price $\pi^U$ for the upward reserve and $\pi^L$ for the downward reserve.

\subsection{Imbalance settlement}
As our formulation is deterministic, the imbalance of an actor can only be caused either by the technical impossibility of satisfying the outcome of the markets, or by an intentional imbalance.
The purpose of this stage is to compute the tariff of upward and downward imbalances, $\pi^{I+}_t$ and $\pi^{I-}_t$.
This tariff is the activation price of the most expensive activated bid.
In case of no imbalance, the imbalance tariff is set to 0. If the imbalance is greater than the contracted reserve, we assume that the SO may use non-contracted reserves to restore balance.
This reserve is supposed to be very expensive and drives the imbalance tariff to the price cap $\pi^{nc}$.

For each period $t$, every actor forecasts the imbalance tariffs $\fc{\pi}^{I^+}_t$ and $\fc{\pi}^{I^-}_t$. This forecast is obtained by the exponential mean of the prices in the last $T$ rounds.
Note that if the imbalance tariff is equal to either zero or $\pi^{nc}$ for an iteration in the history, the value is replaced in the mean by the last one which is not zero or $\pi^{nc}$. 
The models of the actors explicitly take into account these cases.

\subsection{Retailer model}
\label{sec:retailer-model}
In this setting, a retailer estimates the consumption of its clients and make bids at the upper cap price $\pi^{cap}$. 
We assume that a proportion of the retailer's load is flexible and that the retailer has the power to decide when these loads will consume power.
The inelastic part of the demand of the retailer in period $t$ is denoted by $\nu_t$.
We assume that each flexible load $i$ can be accurately represented by a tank model, as in \cite{mathieu2013}.
At each period $t$, the load consumes power $d_{i,t}$ bounded by \eqref{eq:powerLimitsA0}. 
The limits on the energy in the tank are given by \eqref{eq:energyLimitsA0}.
The state transition is given by \eqref{eq:energyBalanceA0}, where $\eta_i$ is the efficiency, $\phi_{i,t}$ the losses in one period, and $\Delta t$ the duration of a period.
The total energy consumed in the time horizon is bounded by \eqref{eq:energyConsumedA0}.
One day ahead, the retailer optimizes the consumption to minimize its retailing costs:
\vspace{-0.5em}
\begin{multline} \label{eq:objA0}
\min \sum_{t=1}^{T} [ \fc{\pi}^E_t D_t + (\pi^{cap}-\fc{\pi}^E_t) \max \{0, D_t-D^{\max}_t \}
\\+ \fc{\pi}^{I+}_t I^+_t + (\pi^{nc}-\fc{\pi}^{I+}) \max \{0, I^+_t-I^{+\max}_t\}
\\+ \fc{\pi}^{I-}_t I^-_t +(\pi^{nc}-\fc{\pi}^{I-}) \max \{0, I^-_t-I^{-\max}_t\} ]
\end{multline}
subject to, $\forall i \in \mathcal{M}, t \in \{1,...,T\}$,
\begin{align}
d^{\min}_{i,t} \leq d_{i,t}, \leq d^{\max}_{i,t} \label{eq:powerLimitsA0}\\
e^{\min}_{i,t} \leq e_{i,t} \leq e^{\max}_{i,t} \label{eq:energyLimitsA0}\\
e_{i,t+1} = e_{i,t} - \phi_{i,t} + \eta_i d_{i,t} \Delta t \label{eq:energyBalanceA0}
\end{align}
$\forall i \in \mathcal{M}$,
\vspace{-0.75em}
\begin{align}
\xi^{\min}_i \leq \sum_{t=1}^{T} d_{i,t} \Delta t \leq \xi^{\max}_i \label{eq:energyConsumedA0}
\end{align}
$\forall t \in \{1,...,T\}$,
\vspace{-0.75em}
\begin{align}
D_t - I^+_t + I^-_t = \nu_t + \sum_{i=1}^M d&_{i,t}. \label{eq:totalPowerA0}
\end{align}

Equation \eqref{eq:totalPowerA0} computes the total demand the retailer submits to the energy market, $D_t$, and the upward (resp. downward) imbalance, $I^+_t$ (resp. $I^-_t) \geq 0$.
From the history of the previous rounds, the retailer learns a threshold demand $D^{\max}_t$ above which the clearing of the energy market yields the price cap $\pi^{cap}$. If in a previous round $\pi^E_t=\pi^{cap}$, then the retailer sets its threshold demand slightly below the volume submitted, e.g., $D^{\max}_t=0.95 D_t$.
The same consideration is applied to imbalance if the tariff of imbalance is either $0$ or $\pi^{nc}$.

After the clearing of the energy market, the retailer runs again the previous optimization problem with $D_t$ given as data to optimize its position for the second and the third stages.

\subsection{Producer model}
\label{sec:producerModel}

The producer optimizes its position using price forecasts and the characteristics
of its production units.
The output of the following optimization problem is the power to be submitted to the energy market $P_t$, the upward/downward reserve $U_t$/$L_t$, and its upward/downward imbalance $I^+_t$/$I^-_t$.
\begin{multline}\label{eq:objProducer}
\max \sum_{t=1}^T [\fc{\pi}^E_t P_t - (\pi^{cap}+\fc{\pi}^E_t) \max \{ 0, P^{\min}_t-P_t \} 
\\- \fc{\pi}^{I+}_t I^+_t - (\pi^{nc}-\fc{\pi}^{I+}) \max \{0, I^+_t-I^{+\max}_t\}
\\- \fc{\pi}^{I-}_t I^-_t - (\pi^{nc}-\fc{\pi}^{I-}) \max \{0, I^-_t-I^{-\max}_t\} 
\\+ \eta_t U_t + \eta_t L_t - \sum_{i} c_{i,t} p_{i,t} ]
\end{multline}
subject to 
$\forall t \in \{1,...,T\}$ and each production unit $i$,
\begin{align}
& p_{i,t}+u_{i,t} \leq p^{\max}_{i,t} &\\
& p^{\min}_{i,t} \leq p_{i,t}-l_{i,t} &\\
& (p_{i,t}+u_{i,t})-p_{i,t-1} \leq \rho^u_i &\\
& p_{i,t-1}-(p_{i,t}-l_{i,t}) \leq \rho^d_i &\\
& u_{i,t}, l_{i,t} \geq 0 &
\end{align}
$\forall t \in \{1,...,T\}$:
\begin{align}
& P_t+I^+_t-I^-_t = \sum_i p_{i,t} &\\
& U_t = \sum_i u_{i,t} &\\
& L_t = \sum_i l_{i,t} &
\end{align}
The power production of unit $i$ is offered to the energy market ($p_{i,t}$) and the reserve market ($u_{i,t}$ as upward reserve, $l_{i,t}$ as downward reserve) at the  price $c_{i,t}$ for every period $t$. 
The predicted downward imbalance is submitted as a bid to the energy market at the price $\fc{\pi}^{I-}$.

In the second stage, $P_t$, the power cleared in the energy market, is a parameter. 
At the third stage, the accepted upward and downward reserve quantities are also fixed, but the producer may still optimize its imbalance.

Similarly to the retailer model, the producer model uses the data from the rounds where the price cap was reached to learn what minimal quantity to submit to the energy market $P^{\min}_t$. The same mechanism is used for the imbalance volume. 
The objective function \eqref{eq:objProducer} uses a price $\epsilon$ to balance the provision of energy and reserve. 
If we set $\eta_t=\pi^U \text{ or } \pi^L$, the previous optimization problem assumes that all reserves proposed by the producer will be accepted at the capacity price. 
To prevent the production of the energy cleared in the energy market at a higher price than the MCP, we put $\eta_t=0.005$\euro/MWh$, \forall t$. 
Other options may be considered, such as taking the product of the capacity prices and the probability of the reserve bid's being accepted.

\subsection{Results}
We simulate a benchmark system for one day, divided into $24$ periods.
This will be used for comparison with the proposal of Section \ref{sec:market2}. 
Producers own two types of units: (i) slow ramping units with costs randomly generated between $45$ and $60$\euro/MWh and (ii) high ramping units with costs between $60$ and $80$\euro/MWh. 
For each period $t$, the SO contracts for a quantity of upward and downward reserves ($R^+_t$ and $R^-_t)$ equal to $2\%$ of the total consumption cleared in MW in the energy market.
The capacity price for the reservation of a reserve is set to $45$\euro/MWh.
The system evolves until the energy prices and the forecasts converge individually and to the same value, or 
a cycle is detected, i.e., actors start taking the same set of positions over and over.

\begin{figure}[tb]
 \centering
 \subfloat [Evolution of the forecasting error (\euro/MWh) for the energy prices ($\|\fc{\pi}^E-\pi^E\|_{\infty}$) for a system with $6\%$ of flexible consumption.]{\label{fig:article0-006-forecastBias}\includegraphics[height=4cm]{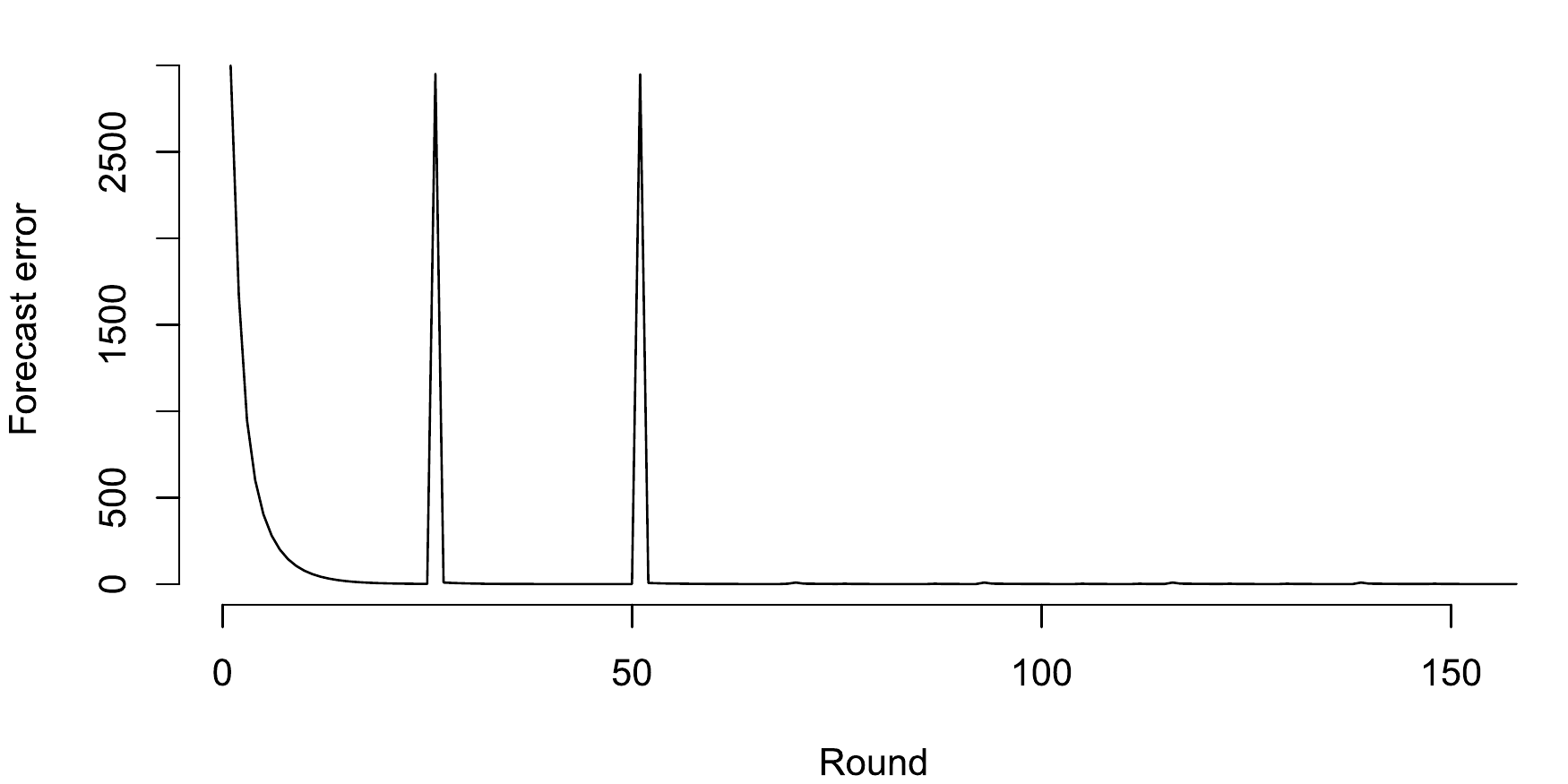}}\\
 \subfloat[Variability of the energy price (\euro/MWh) as a function of the amount of flexible consumption.]{\label{fig:EPV}\includegraphics[height=4cm]{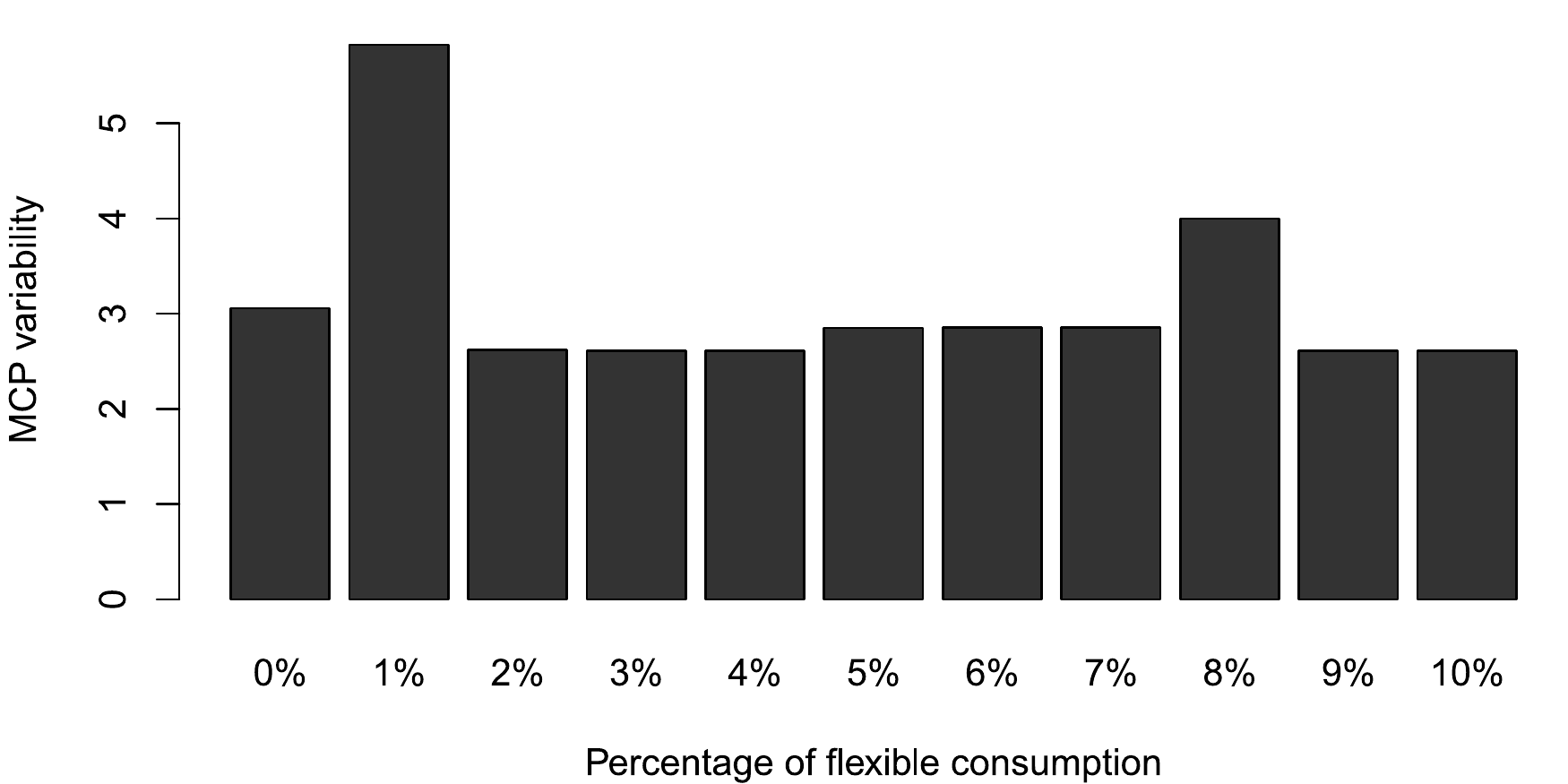}} \\
 \subfloat [Mean total imbalance (MWh) as a function of the amount of flexible consumption.]{\label{fig:TI} \includegraphics[height=4cm]{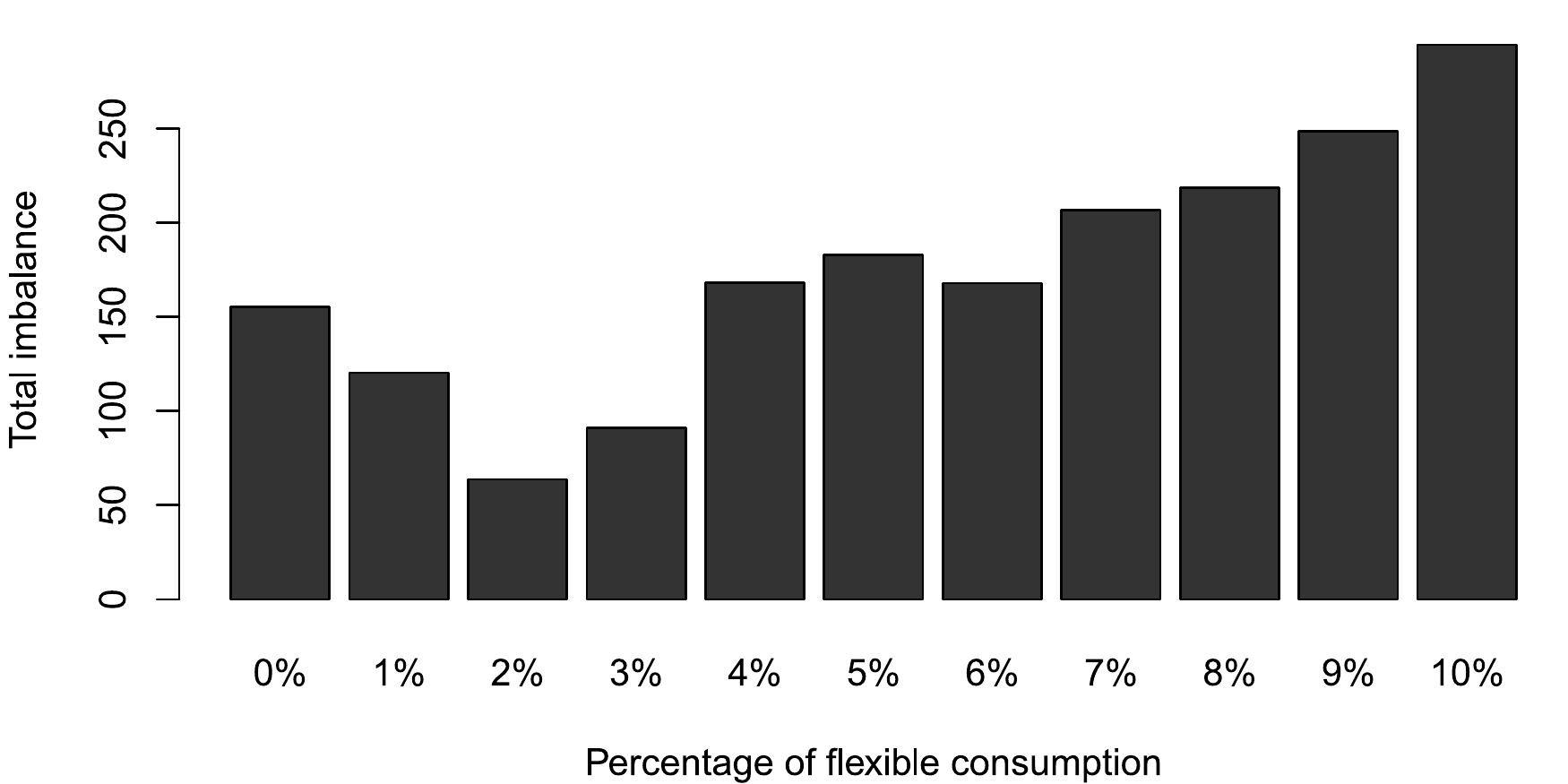}} 
\caption{Simulation of the current system.}
\end{figure}

We present a typical run for a mean total consumption of $1000$ MW with $6\%$ of flexible loads.
Fig. \ref{fig:article0-006-forecastBias} shows for each round a measure of the MCP forecasting error.
The system cycles after $113$ rounds, repeating rounds $77$ to $112$.
The following results are the mean over these 35 rounds. 
The mean energy market price is $49.81$\euro/MWh.
The SO cost for reserve procurement is $42683$\euro.
This reserve covers a total imbalance of $190$ MWh over the day and no non-contracted reserve needs to be used. 
We observe that this imbalance is caused solely by the producers.

We now compare the status of the system for a mean total consumption of $1000$ MW with $0\%$ to $10\%$ of flexible loads.
With our parameters, increasing the flexibility of the demand-side left the mean energy market price and SO costs for reserve procurement barely unchanged.
The variability of the energy price, defined here as the difference between the minimum and maximum price, is given in Fig. \ref{fig:EPV}.
Fig. \ref{fig:TI} shows the mean total imbalance of the system as a function of the amount of flexible loads.
Non-contracted reserve is not needed in any of the cases considered.

\section{Opening the reserve market to retailers}
\label{sec:market2}
We focus here on the provision of secondary reserves by retailers using load flexibility.
First, we introduce modulation bids for the reserve market that are suitable to demand side management.
Sections \ref{sec:modelReserveMarket}, \ref{sec:modelImbalanceSettlement} and \ref{sec:retailerModel2} propose, respectively, the required modifications to the models of the reserve market, the imbalance settlement, and the retailer.
Section \ref{sec:results2} presents the result of opening the reserve market to retailers.

\subsection{Modulation bids for the reserve market}
Unlike production units, the consumption of a load in a period depends on the consumption in the previous periods. 
Increasing the consumption in one period implies that the consumption will decrease later on.
This fact motivates the introduction of bids more adapted to load behavior.
A modulation bid consists of a flexibility margin $[D_t-F,D_t+F] \ \forall t \in \mathcal{N}$ around a baseline consumption $D_t$ over a set $\mathcal{N}$ of consecutive market periods. $F$ is the maximum amplitude of the power modulation.
The SO can specify the consumption of the retailer for every period $t \in \mathcal{N}$ under the constraints that these specifications do not violate the margin and that the total energy consumed in the $\mathcal{N}$ periods is identical.
The total energy consumed must be constant in order to provide the same utility to the load owners. For instance, a heat pump turned off for an hour needs to consume more afterwards to get the temperature back to its set point.

\subsection{Clearing of the reserve market}
\label{sec:modelReserveMarket}
The following optimization model considers the cost of reservation as well as the cost of activation $c^E_i$ of each bid.
The  capacity price of modulation bids is regulated at $\pi^F$. 
As the SO has no knowledge of the future imbalance, it supposes that it may activate all the contracted reserves.
The objective function \eqref{eq:objReserveClearing} shows that the SO receives the activation cost of downward reserve bids.
To prevent this model from contracting every downward reserve bid, we introduce an additional modeling parameter $c^o_t$ that penalizes the amount of reserve contracted over the requirements $R^+_t$ and $R^-_t$ in period $t$.
Our implementation uses $c^o_t = 1.1 \max_{i \in \mathcal{P}^-_t} c^E_i$.
The efficiency factor $\zeta$ of a modulation bid expresses the fact that modulation bids are not equivalent to classical bids.

\subsubsection{Data}
\begin{tabbing}
\hspace{1cm}\=\hspace{1cm}\=\kill
$R^+_t$ \> Quantity of upward reserve to contract for period $t$.\\
$R^-_t$ \> Quantity of downward reserve to contract for period $t$.\\
$\mathcal{P}^+_t$ \> Set of classical upward bids for period $t$.\\
$\mathcal{P}^-_t$ \> Set of classical downward bids for period $t$.\\
$\mathcal{F}$ \> Set of modulation bids.\\
$\mathcal{B}$ \> Set of bids : $\mathcal{B}=\mathcal{F} \cup_{t=1}^T (\mathcal{P}^+_t \cup \mathcal{P}^-_t)$\\
$Q_i$ \> Volume of bid $i \in \mathcal{B}$.\\
$c^E_i$ \> Cost of activation of bid $i \in \mathcal{B}$.\\
$\mathcal{N}_i$ \> Set of periods covered by the modulation bid $i \in \mathcal{F}$.\\
$\zeta_i$ \> Efficiency ratio of bid $i \in \mathcal{B}$.\\
$\pi^{nc}$ \> Cost of activating non-contracted reserve.
\end{tabbing}

\subsubsection{Variables}
\begin{tabbing}
\hspace{1cm}\=\hspace{1cm}\=\kill
$x_i$ \> Determine if a bid $i \in \mathcal{B}$ is accepted totally ($=1$),\\
\> partially ($\in ]0,1[$) or rejected ($=0$).\\
$s^+_t$ \> Over-contracted upward reserve: $s^+_t \geq 0$.\\
$s^-_t$ \> Over-contracted downward reserve: $s^-_t \geq 0$.\\
$n^+_t$ \> Non-contracted upward reserve: $s^+_t \geq 0$.\\
$n^-_t$ \> Non-contracted downward reserve: $s^-_t \geq 0$.
\end{tabbing}

\subsubsection{Model}
\vspace{-0.5em}
\begin{multline}\label{eq:objReserveClearing}
\min \sum_{t=1}^{T} \Bigg[\sum_{i \in \mathcal{P}^+_t} (\pi^U+c^E_i) x_i Q_i + \sum_{i \in \mathcal{P}^-_t} (\pi^L-c^E_i) x_i Q_i \\+ c^o_t (s^+_t + s^-_t) + \pi^{nc} (n^+_t + n^-_t)\Bigg] +\sum_{i \in \mathcal{F}} (\pi^F+c^E_i) x_i Q_i
\end{multline}
subject to
$\forall t \in \{1,...,T\}$,
\begin{align}
\sum_{i \in \mathcal{P}^+_t} Q_i x_i \zeta_i + \sum_{i \in \mathcal{F}: t \in \mathcal{N}_i} Q_i x_i \zeta_i + n^+_t - s^+_t= R^+_t \\
\sum_{i \in \mathcal{P}^-_t} Q_i x_i \zeta_i + \sum_{i \in \mathcal{F}: t \in \mathcal{N}_i} Q_i x_i \zeta_i + n^-_t - s^-_t = R^-_t
\end{align}

\subsection{Imbalance settlement}
\label{sec:modelImbalanceSettlement}
The following model gives the optimal activation scheme a SO would use to restore balance in every market period.
We assume the SO knows exactly what the imbalance $I_t$ will be for each period, given the nominations of the actors, i.e. their positions on the day-ahead energy market.
\subsubsection{Data}
\begin{tabbing}
\hspace{1cm}\=\hspace{1cm}\=\kill
$I_t$ \> Imbalance in period $t$.\\
$\mathcal{P}^+_t$ \> Set of contracted upward reserve bids for period $t$.\\
$\mathcal{P}^-_t$ \> Set of contracted downward reserve bids for period $t$.\\
$\mathcal{F}$ \> Set of contracted modulation bids.\\
$Q_i$ \> Volume of bid $i \in \mathcal{B}_t, \ \forall t$.\\
$c^E_i$ \> Cost of usage of bid $i \in \mathcal{B}$.\\
$\mathcal{N}_i$ \> Set of periods covered by the modulation bid $i \in \mathcal{F}$.\\
$\pi^{nc}$ \> Cost of activating non-contracted reserve.
\end{tabbing}

\subsubsection{Variables}
\begin{tabbing}
\hspace{1cm}\=\hspace{1cm}\=\kill
$x_i$ \> Activation of a bid $i \in \mathcal{P}^+_t \cup \mathcal{P}^-_t,\ \forall t$: $x_i \in [0,1]$.\\
$v_{i,t}$ \> Activation upward of a modulation bid $i \in \mathcal{F}$\\
\>in period $t \in \mathcal{N}_i$: $v_{i,t} \in [0,1]$.\\
$w_{i,t}$ \> Activation downward of a modulation bid $i \in \mathcal{F}$\\
\>in period $t \in \mathcal{N}_i$: $w_{i,t} \in [0,1]$.\\
$y^+_t$ \> Non-contracted upward reserve: $y^+_t \geq 0$.\\
$y^-_t$ \> Non-contracted downward reserve: $y^-_t \geq 0$.
\end{tabbing}

\subsubsection{Model}
\begin{multline}
\min \sum_{i \in \mathcal{F}} c^E_i Q_i \sum_{\tau \in \mathcal{N}_i} (v_{i,\tau}+w_{i,\tau}) \\+ \sum_{t=1}^{T} \left[\sum_{i \in \mathcal{P}^+_t} c^E_i x_i Q_i + \sum_{i \in \mathcal{P}^-_t} (c^o_t-c^E_i) x_i Q_i + \pi^{nc} (y^+_t + y^-_t) \right]
\end{multline}
subject to $\forall t \in \{1,...,T\}$,
\begin{multline}
\sum_{i \in \mathcal{P}^+_t} Q_i x_i - \sum_{i \in \mathcal{P}^-_t} Q_i x_i+ \sum_{i \in \mathcal{F}: t \in \mathcal{N}_i} Q_i (v_{i,t}-w_{i,t} )
\\+ y^+_t - y^-_t + I_t = 0
\end{multline}
$\forall i \in \mathcal{F}$, 
\begin{align} \label{eq:integralityConstraint}
\sum_{t \in \mathcal{N}_i} (v_{i,t}-w_{i,t})=0
\end{align}

\subsection{Retailer model to provide secondary reserve}
\label{sec:retailerModel2}
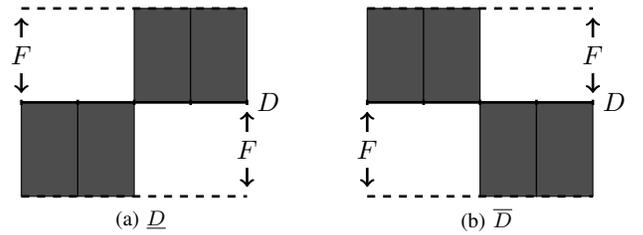
\begin{figure}[h]
\begin{tabular}{cc}
\subfloat[$\ls{D}$]{
	\definecolor{darkGrey}{RGB}{77,77,77}
	\begin{tikzpicture}[x=1.5cm,y=2.5cm]
		\filldraw[draw=black,fill=darkGrey] (0,0) rectangle (.5,-.5);
		\filldraw[draw=black,fill=darkGrey] (.5,0) rectangle (1,-.5);
		\filldraw[draw=black,fill=darkGrey] (1,0) rectangle (1.5,.5);
		\filldraw[draw=black,fill=darkGrey] (1.5,0) rectangle (2,.5);
		\draw[-,>=latex, line width=1pt](0,0)--(2,0) node[right] {$D$};
	 \foreach \x in {0,...,4}
	 	\draw[line width=1pt] (\x/2,1pt) -- (\x/2,-1pt);
	 \draw[dashed,line width=1pt](0,.5)--(2,.5);
	 \draw[dashed,line width=1pt](0,-.5)--(2,-.5);
	 \draw[<->,line width=1pt] (0,.05) -- (0,.45) node[midway,fill=white] {$F$}; 
	 \draw[<->,line width=1pt] (2,-.05) -- (2,-.45) node[midway,fill=white] {$F$}; 
	\end{tikzpicture}
	\label{fig:bidPeriodPower-underline}
} 
&
\subfloat[$\us{D}$]{
	\definecolor{darkGrey}{RGB}{77,77,77}
	\begin{tikzpicture}[x=1.5cm,y=2.5cm]
		\filldraw[draw=black,fill=darkGrey] (0,0) rectangle (.5,.5);
		\filldraw[draw=black,fill=darkGrey] (.5,0) rectangle (1,.5);
		\filldraw[draw=black,fill=darkGrey] (1,0) rectangle (1.5,-.5);
		\filldraw[draw=black,fill=darkGrey] (1.5,0) rectangle (2,-.5);
		\draw[-,>=latex, line width=1pt](0,0)--(2,0) node[right] {$D$};
	 \foreach \x in {0,...,4}
	 	\draw[line width=1pt] (\x/2,1pt) -- (\x/2,-1pt);
	 \draw[dashed,line width=1pt](0,.5)--(2,.5);
	 \draw[dashed,line width=1pt](0,-.5)--(2,-.5);
	 \draw[<->,line width=1pt] (0,-.05) -- (0,-.45) node[midway,fill=white] {$F$}; 
	 \draw[<->,line width=1pt] (2,.05) -- (2,.45) node[midway,fill=white] {$F$}; 
	\end{tikzpicture}
} 
\end{tabular}
\caption{Illustration of the two most constraining scenarios for the provision of a modulation.}
\label{fig:bidPeriodPower}
\end{figure}

The objective of the retailer is to maximize its profit from the retailing activities and the flexibility services it sells to the SO.
We suppose that the retailer selects a set of modulation bids $\mathcal{R}$ to submit to the reserve market, whose quantity is the result of the following optimization problem. These bids are supposed to be non-overlapping. Each bid $k$ starts in period $\tau_k$ and lasts $N_k$ periods.
For every modulation bid $k$, we define the two most constraining scenarios for the provision of the modulation range $[D_t-F_k,D_t+F_k] \ \forall t \in \mathcal{N}_k$: $\us{D}$ and $\ls{D}$ (cf. Fig.~\ref{fig:bidPeriodPower}) where the notation underline/overline indicates the variables related to these scenarios.
In the first one, $\us{D}$, the SO asks for a modulation upwards for the $N/2$ first periods and ensures an energy balance in the last two periods.
In the second one, $\ls{D}$, the SO asks for modulation downwards for the $N/2$ first periods. 
The two scenarios $\us{D}$ and $\ls{D}$ may be used to define the range of flexibility described previously. 
Section \ref{sec:proofScenarios} proves that scenarios $\us{D}$ and $\ls{D}$ cover every activation scheme that the SO may ask for from a retailer within the limits $[D_t-F,D_t+F] \ \forall t \in \mathcal{N}$ if loads are modeled by \eqref{eq:powerLimitsA}--\eqref{eq:energyBalanceA}.
The cost of activation of the modulation bids of the retailer is considered to be null, as the utility of the load is ensured by the integrality constraint of the bid \eqref{eq:integralityConstraint}.

The optimization problem solved by the retailer is:
\begin{multline} \label{eq:objAA}
\min \sum_{t=1}^{T} [ \fc{\pi}^E_t D_t + (\pi^{cap}-\fc{\pi}^E_t) \max \{0, D_t+I^-_t-D^{\max}_t \}
\\+ \fc{\pi}^{I+}_t I^+_t + (\pi^{nc}-\fc{\pi}^{I+}) \max \{0, I^+_t-I^{+\max}_t\}
\\+ \fc{\pi}^{I-}_t I^-_t +(\pi^{nc}-\fc{\pi}^{I-}) \max \{0, I^-_t-I^{-\max}_t\}
\\- \sum_{k \in \mathcal{R}} \left(\sum_{t=\tau_k}^{\tau_k+N-1} \pi^F_t \right) F_k ]
\end{multline}
subject to
$\forall t \in \{1,...,T\}$,
\begin{align}
D_t-I^+_t+I^-_t = \nu_t + \sum_{i=1}^M d_{i,t} \label{eq:totalPowerA}\\
\us{D}_t = \nu_t + \sum_{i=1}^M \us{d}_{i,t} \label{eq:totalPowerUpA}\\
\ls{D}_t = \nu_t + \sum_{i=1}^M \ls{d}_{i,t} \label{eq:totalPowerDownA}
\end{align}
$\forall i \in \mathcal{M}, t \in \{1,...,T\}$,
\begin{align}
d^{\min}_{i,t} \leq d_{i,t}, \us{d}_{i,t}, \ls{d}_{i,t} \leq d^{\max}_{i,t} \label{eq:powerLimitsA}\\
e^{\min}_{i,t} \leq e_{i,t}, \us{e}_{i,t}, \ls{e}_{i,t} \leq e^{\max}_{i,t} \label{eq:energyLimitsA}\\
e_{i,t+1} = e_{i,t} - \phi_{i,t} + \eta_i d_{i,t} \Delta t \label{eq:energyBalanceA}
\end{align}
$\forall i \in \mathcal{M}$,
\begin{align}
\xi^{\min}_i \leq \sum_{t=1}^{T} d_{i,t} \Delta t \leq \xi^{\max}_i \label{eq:energyConsumedA}
\end{align}
$\forall i \in \mathcal{M}, k \in \mathcal{R}, t=\tau_k$,
\begin{align}
\us{e}_{i,t+1} = e_{i,t} - \phi_{i,t} + \eta_i \us{d}_{i,t} \Delta t \label{eq:energyBalanceUpBeginBidA}\\
\ls{e}_{i,t+1} = e_{i,t} - \phi_{i,t} + \eta_i \ls{d}_{i,t} \Delta t \label{eq:energyBalanceDownBeginBidA}
\end{align}
$\forall i \in \mathcal{M}, k \in \mathcal{R}, t \in \{\tau_k+1,...,\tau_k+N_k-2\}$,
\begin{align}
\us{e}_{i,t+1} = \us{e}_{i,t} - \phi_{i,t} + \eta_i \us{d}_{i,t} \Delta t \label{eq:energyBalanceUpIntraBidA}\\
\ls{e}_{i,t+1} = \ls{e}_{i,t} - \phi_{i,t} + \eta_i \ls{d}_{i,t} \Delta t \label{eq:energyBalanceDownIntraBidA}
\end{align}
$\forall i \in \mathcal{M}, k \in \mathcal{R}, t=\tau_k+N_k-1$,
\begin{align}
e_{i,t+1} = \us{e}_{i,t} - \phi_{i,t} + \eta_i \us{d}_{i,t} \Delta t \label{eq:energyBalanceUpEndBidA}\\
e_{i,t+1} = \ls{e}_{i,t} - \phi_{i,t} + \eta_i \ls{d}_{i,t} \Delta t \label{eq:energyBalanceDownEndBidA}
\end{align}
$\forall k \in \mathcal{R}, t \in [\tau_k,\tau_k+N_k/2-1]$,
\begin{align}
F_k \leq \us{D}_t-(D_t-I^+_t+I^-_t) \label{eq:flexUpStartA}\\
F_k \leq (D_t-I^+_t+I^-_t)-\ls{D}_t \label{eq:flexDownStartA}
\end{align}
$\forall k \in \mathcal{R}, t \in [\tau_k+N_k/2,\tau_k+N_k-1]$,
\begin{align}
F_k \leq \ls{D}_t-(D_t-I^+_t+I^-_t) \label{eq:flexUpEndA}\\
F_k \leq (D_t-I^+_t+I^-_t)-\us{D}_t \label{eq:flexDownEndA}
\end{align}

The energy of each load for the modulation scenario $\us{D}, \ls{D}$ is given by \eqref{eq:energyBalanceUpBeginBidA} and \eqref{eq:energyBalanceDownBeginBidA} at the beginning of each bid period, by \eqref{eq:energyBalanceUpIntraBidA} and \eqref{eq:energyBalanceDownIntraBidA} in the middle, and by \eqref{eq:energyBalanceUpEndBidA} and \eqref{eq:energyBalanceDownEndBidA} at the end. 
The available volumes of modulation are given by \eqref{eq:flexUpStartA}--\eqref{eq:flexDownEndA}.

\subsection{Results}
\label{sec:results2}
\begin{figure}[tb]
 \centering
 \subfloat [Total imbalance (MWh).]{\label{fig:TI_2}\includegraphics[height=4cm]{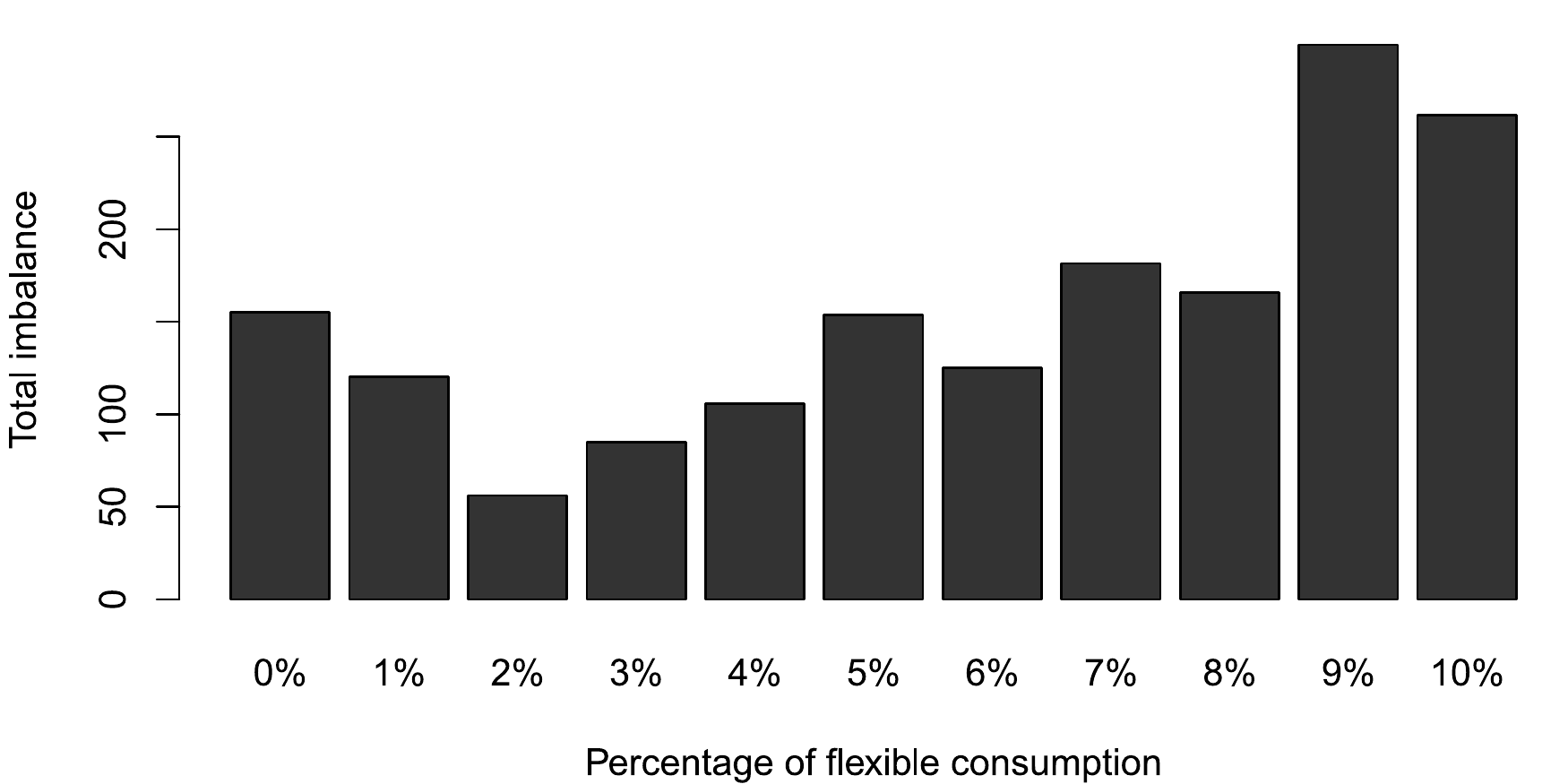}} \\
 \subfloat [SO reserve procurement cost (k\euro).]{\label{fig:TSO_RPC}\includegraphics[height=4cm]{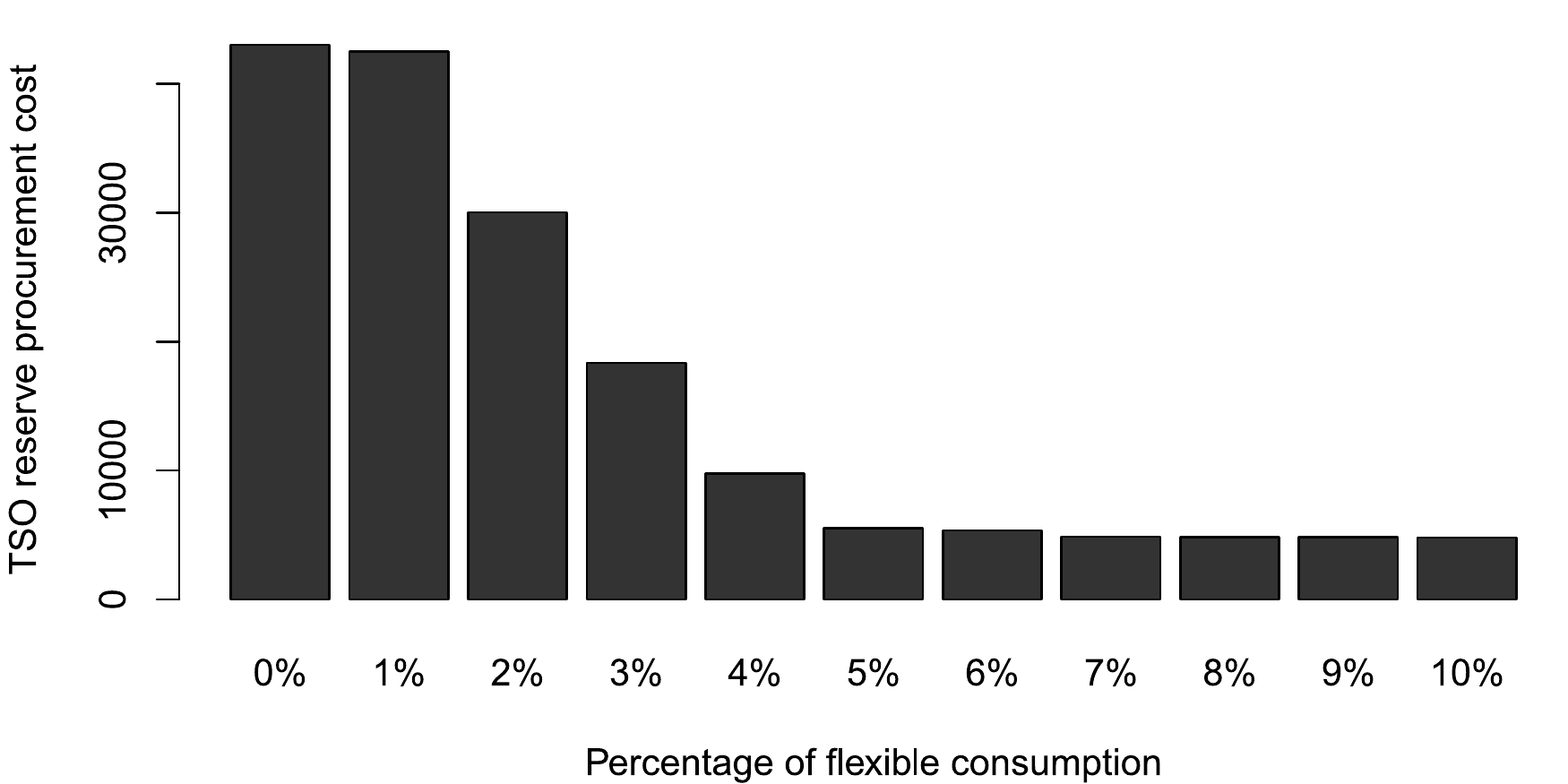}} \\
 \subfloat [Volume of non-contracted reserves (MWh) used.]{\label{fig:TSO_NCR} \includegraphics[height=4cm]{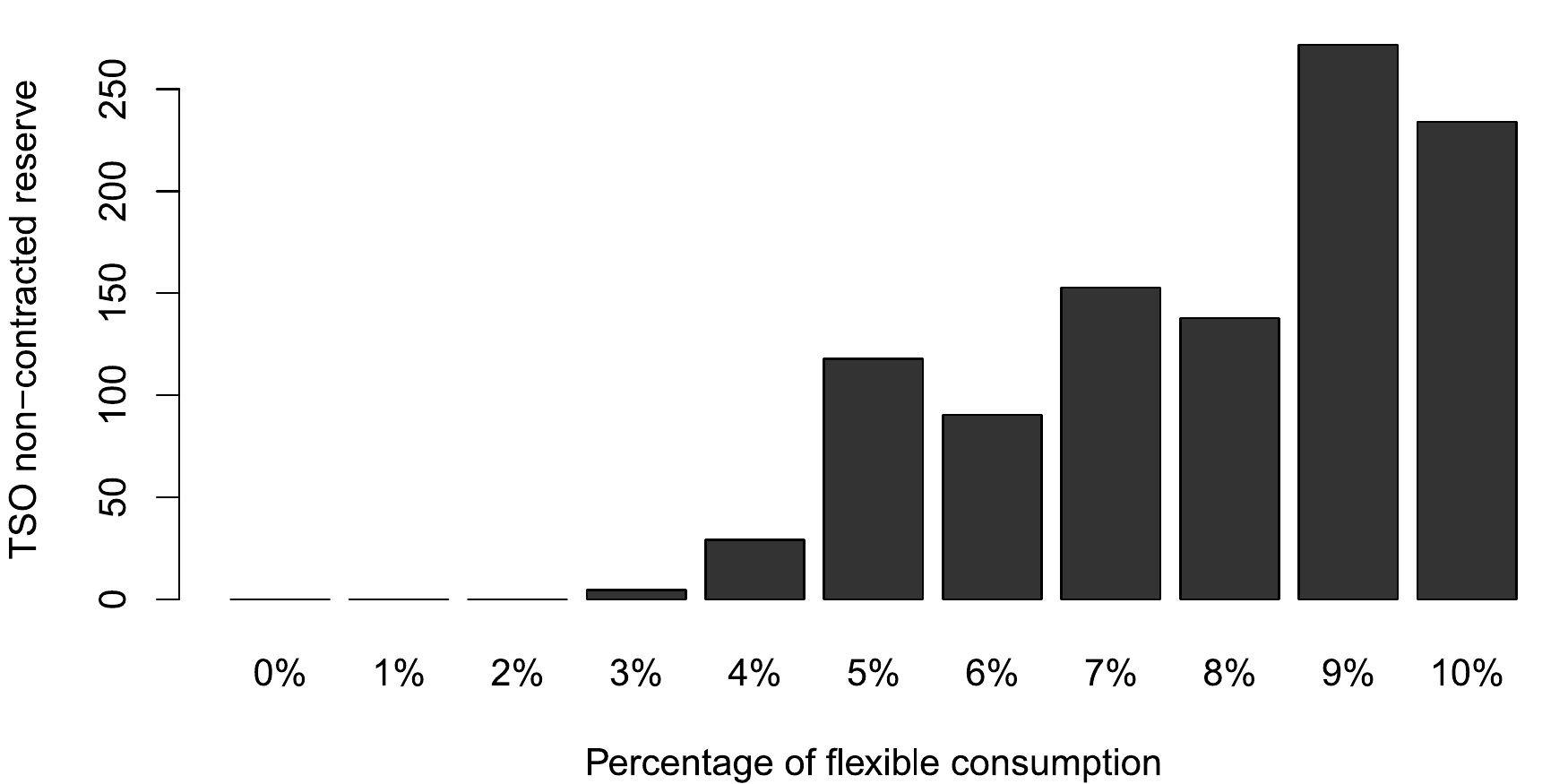}} 
\caption{Impact of load aggregation included as modulation bids in the reserve market, as a function of the flexibility rate.}
\end{figure}

We now run the system with a reserve market open to modulation bids. 
The reservation price of these bids is set to $10$\euro/MWh.
Retailers submit a modulation bid for every four hours.
We use an efficiency factor of $0.5$ to express the fact that the quantity brought by modulation bids is worth one-half that of the quantity from a classical reserve bid.
The results are reported for flexibility rates between $0$ and $10\%$.
Once again, the mean energy market price is left barely unchanged, at around $49.81$\euro/MWh. 
The price variability is similar to that observed in Fig. \ref{fig:EPV}.
The total imbalance, shown in Fig. \ref{fig:TI_2}, is slightly lower than in Fig. \ref{fig:TI}.
Fig. \ref{fig:TSO_RPC} shows that, with flexibility, the SO reserve procurement costs decrease significantly, to only $11\%$ of the initial costs.
Nevertheless, the volume of non-contracted reserves that have to be used is increasing in the flexibility rate, as shown by Fig. \ref{fig:TSO_NCR}.
This volume of non-contracted reserve is mostly due to the inability of a modulation bid to sustain an imbalance of the same sign for its whole time horizon.

\section{Conclusion}
\label{sec:conclusion}
An agent-based model has been introduced to study the introduction of load aggregation in the secondary reserve market.
In this model, each actor maximizes its profit based on a forecast of the prices.
Producers and retailers (which perform load aggregation) optimize their positions in the energy and reserve markets, and in the settlement of imbalances. 
We propose to add a new product, the modulation bid, to the reserve market, that takes into account the inter-dependency between time periods due to load constraints. 
The results show that introducing this product decreases drastically the cost for reserve procurement. 
Unfortunately, modulation bids are not efficient at covering an imbalance of the same sign for multiple periods, which results in the activation of non-contracted reserves. 
This can be avoided by contracting for more reserves.

%
The proposed agent-based model could be modified to assess variants of the market model studied in this paper.
One variant that would be worth exploring is to include energy constrained bids in the day-ahead energy market \cite{Su20091199}.
One could also consider market-based capacity prices for the reserve.
They could decrease the cost of reserve procurement but may, however, lead to gaming.
Finally, an extension to the provision of services to solve congestion or over-voltage problems in the distribution network should be investigated.

\section*{Acknowledgments}
This research was supported by the public service of Wallonia – Department of Energy and Sustainable Building, within the framework of the GREDOR project.
The authors are grateful for the financial support of the Belgian Network DYSCO, an Inter-University Attraction Poles Program initiated by the Science Policy Office of the government of Belgium.

\section{Appendix}
\label{sec:proofScenarios}
We prove here that scenarios $\us{D}$ and $\ls{D}$ cover every activation scheme within the limits $[D_t-F,D_t+F] \ \forall t \in \mathcal{N}$ for which the SO may ask if loads are modeled by \eqref{eq:powerLimitsA}--\eqref{eq:energyBalanceA}.
For ease of exposition, we develop our argument on the set of periods $\mathcal{N}=\{1,...,N\}$. 
We study the behavior of each load individually and drop the load index for conciseness. 
Every scenario * must satisfy energy balance:
\begin{align}
& \starred{e}_2=e_1 - \phi_1 + \eta \starred{d}_1 \Delta t & \label{eq:firstEnergy}\\
& \starred{e}_{t+1}=e_t - \phi_t + \eta \starred{d}_t \Delta t & \forall t \in [2,N-1] \label{eq:intraBidEnergy}\\
& e_{N+1} = e_N - \phi_N + \eta \starred{d}_N \Delta t. & \label{eq:localEnergyBalance}
\end{align}

We use $'$ to refer to a random scenario. $d'$ obeys the constraints \eqref{eq:firstEnergy}--\eqref{eq:localEnergyBalance}. If we use \eqref{eq:intraBidEnergy} and \eqref{eq:firstEnergy} in \eqref{eq:localEnergyBalance}, we get
\begin{equation}
e_{N+1}=e_1 - \sum_{t=1}^N \phi_t + \eta \Delta t \sum_{t=1}^N \starred{d}_t
\end{equation}
which is true for every $\starred{d}$ and in particular for $\starred{d}=d$ and $\starred{d}=d'$. Therefore, we can identify this equality with the total energy consumed in the bid period:
\begin{equation}
\sum_{t=1}^N d'_t = \sum_{t=1}^N d_t \label{eq:powerConsumedBid}
\end{equation}

We want to prove that every load scenario $d': d'_t \in [\ls{d}_t, \us{d}_t] \ \forall t$ satisfying \eqref{eq:firstEnergy}--\eqref{eq:localEnergyBalance} is feasible if $d$, $\ls{d}$ and $\us{d}$ are feasible.
If we use \eqref{eq:intraBidEnergy} and \eqref{eq:firstEnergy}, we have
\begin{align}
& \starred{e}_{t+1}=e_1 - \sum_{\tau=1}^t \phi_{\tau} + \eta \Delta t \sum_{\tau=1}^t \starred{d}_{\tau} & \forall t \in [2,N]. \label{eq:progressiveEnergy}
\end{align}
As $\ls{d}$ and $\us{d}$ are feasible and \eqref{eq:powerConsumedBid} holds, we have for the first one-half of the bid period $[1,N/2]$:
\begin{align}
& d^{\min}_t \leq \ls{d}_t \leq d'_t \leq \us{d}_t \leq d^{\max}_t & \forall t \in [1,N/2] \label{eq:randomPowerFirstHalf}
\end{align}
and for the next half,
\begin{align}
& d^{\min}_t \leq \us{d}_t \leq d'_t \leq \ls{d}_t \leq d^{\max}_t & \forall t \in [N/2+1,N]. \label{eq:randomPowerSecondHalf}
\end{align}
Employing \eqref{eq:progressiveEnergy} and \eqref{eq:randomPowerFirstHalf} and using the fact that $\ls{d}$ and $\us{d}$ are feasible scenarios, we have
\begin{align}
& e^{\min}_t \leq \ls{e}_t \leq e'_t \leq \us{e}_t \leq e^{\max}_t & \forall t \in [1,N/2] \label{eq:randomEnergyFirstHalf}
\end{align}
and similarly for the remaining periods,
\begin{align}
& e^{\min}_t \leq \us{e}_t \leq e'_t \leq \ls{e}_t \leq e^{\max}_t & \forall t \in [N/2+1,N]. \label{eq:randomEnergySecondHalf}
\end{align}
Now we can see that every scenario $': D'_t \in [\ls{D}_t, \us{D}_t]$ which satisfies \eqref{eq:firstEnergy}--\eqref{eq:localEnergyBalance} is feasible, by summing over the whole portfolio of loads.

\printbibliography

\end{document}